\documentclass[aps, prx,twocolumn,longbibliography,superscriptaddress,amsmath,amssymb,floatfix]{revtex4-2}

\usepackage{amssymb}
\usepackage{graphicx}
\usepackage{amsmath}
\usepackage{color}
\usepackage{mathrsfs}
\usepackage{float}
\usepackage{indentfirst}
\usepackage{mathrsfs}
\usepackage{float}
\usepackage{indentfirst}
\usepackage{textcomp}
\usepackage{comment}
\usepackage{mathtools}
\usepackage{natbib,hyperref}
\usepackage{soul}

\begin{document}
\title{Augmenting Density Matrix Renormalization Group with Clifford Circuits}

\author{Xiangjian Qian}
\affiliation{Key Laboratory of Artificial Structures and Quantum Control (Ministry of Education),  School of Physics and Astronomy, Shanghai Jiao Tong University, Shanghai 200240, China}

\author{Jiale Huang}
\affiliation{Key Laboratory of Artificial Structures and Quantum Control (Ministry of Education),  School of Physics and Astronomy, Shanghai Jiao Tong University, Shanghai 200240, China}

\author{Mingpu Qin} \thanks{qinmingpu@sjtu.edu.cn}
\affiliation{Key Laboratory of Artificial Structures and Quantum Control (Ministry of Education),  School of Physics and Astronomy, Shanghai Jiao Tong University, Shanghai 200240, China}

\affiliation{Hefei National Laboratory, Hefei 230088, China}

\date{\today}


\begin{abstract}

    Density Matrix Renormalization Group (DMRG) is widely acknowledged as a highly effective and accurate method for solving one-dimensional quantum many-body systems. However, the direct application of DMRG to the study of two-dimensional systems encounters challenges due to the limited entanglement encoded in the underlying wave-function ansatz, known as Matrix Product State (MPS). Conversely, Clifford circuits offer a promising avenue for simulating states with substantial entanglement, albeit confined to stabilizer states. In this work, we present the seamless integration of Clifford circuits within the DMRG algorithm, leveraging the advantages of both Clifford circuits and DMRG. This integration leads to a significant enhancement in simulation accuracy with small additional computational cost. Moreover, this framework is useful not only for its current application but also for its potential to be easily adapted to various other numerical approaches.
    
\end{abstract}

\maketitle
{\em Introduction --}
The exploration of strongly correlated quantum many-body systems is a center topic of condensed matter physics, as it frequently unveils exotic quantum states and novel physical phenomena \cite{RevModPhys.66.763,Xiao:803748,bookc,marino_2017,doi:10.1126/science.1091806,senthil2023deconfined}. However, simulating quantum many-body systems presents a central challenge in modern physics, primarily due to the exponential size of the underlying Hilbert space and the intricate quantum correlations involved. To address these challenges, powerful numerical methods become necessary \cite{PhysRevX.5.041041}. Density Matrix Renormalization Group (DMRG) \cite{PhysRevLett.69.2863} represents a powerful numerical framework for analyzing and simulating one-dimensional (1D) quantum many-body systems. The underlying wave-function ansatz of DMRG is known as Matrix Product State (MPS) \cite{PhysRevLett.75.3537}, which provides a succinct yet powerful representation of 1D quantum states. Past investigations have underscored the efficiency of DMRG in studying (quasi) 1D systems \cite{RevModPhys.77.259,SCHOLLWOCK201196}, establishing it as the workhorse for such studies. Nevertheless, directly applying DMRG to the realm of two-dimensional (2D) systems has proven less successful compared to its 1D counterpart. This limitation primarily arises from the constrained entanglement encoded within the underlying wave-function ansatz \cite{PhysRevB.49.9214}.


To address the limitation of limited entanglement in MPS, several new ansatzes have been introduced. Examples include Projected Entangled Pair States (PEPS) for 2D systems \cite{2004cond.mat..7066V,RevModPhys.93.045003,ORUS2014117}, 2D Multiscale Entanglement Renormalization Ansatz (MERA) \cite{PhysRevLett.102.180406,PhysRevLett.99.220405}, Projected Entangled Simplex States (PESS) for 2D systems \cite{PhysRevX.4.011025}, and so on \cite{annurev-conmatphys-040721-022705,xiang2023density,PhysRevLett.97.107206}. However, these methods often come with high computational costs, which hampers the study using large bond-dimensions.

\begin{figure*}[t]
    \includegraphics[width=170mm]{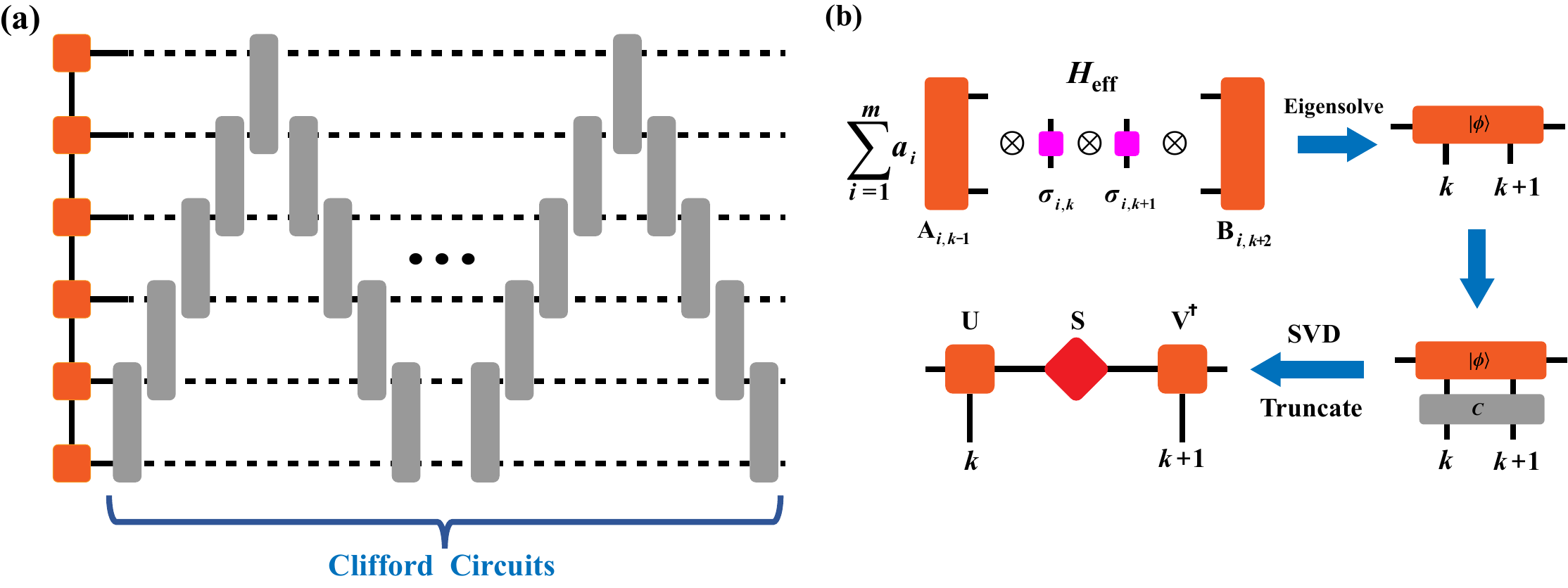}
       \caption{(a) Diagrammatic representation of the wave function for CAMPS. CAMPS enhances the capability of MPS by incorporating additional Clifford circuits on the physical bonds of MPS. (b) Schematic illustration of the optimization procedure in CAMPS. We first calculate the ground state $|\phi\rangle$ of the effective Hamiltonian $H_{\text{eff}}$ shown in Eq.~(\ref{H_eff}) as in DMRG. Instead of directly employing a truncated SVD on $|\phi\rangle$, we apply a two-qubit Clifford circuit $C$ to $|\phi\rangle$ to minimize discarded singular values during truncation. Considering that $C H_{\text{eff}}C^\dagger C|\phi\rangle=E_g C|\phi\rangle$, we then need to perform a transformation to the original Hamiltonian $H^\prime =CHC^\dagger$. The explicit expression for $H^\prime$ is detailed in Eq.~(\ref{H_prime}). Additional details can be found in the main text.}
       \label{CAMPS}
\end{figure*}

Recently, Fully-augmented Matrix Product States (FAMPS) \cite{Qian_2023} were proposed, aiming to strike a balance between the entanglement captured in the ansatz and computational efficiency. FAMPS augments MPS with unitary transformations of physical degrees of freedom, known as disentanglers \cite{PhysRevLett.99.220405,wu2022disentangling}. Notably, FAMPS has been shown to support area-law like entanglement in 2D systems while maintaining the computational efficiency of MPS with small overhead. Numerical investigations have further demonstrated the enhanced accuracy of FAMPS compared to MPS across a range of prominent two-dimensional spin models \cite{Qian_2023, PhysRevB.109.L161103}.

In the field of quantum computing and quantum information \cite{Nielsen_Chuang_2010}, the Gottesman-Knill theorem states that quantum circuits that consist solely of Clifford gates (Hadamard gate, the phase gate $S$, and the controlled-NOT gate) can be efficiently simulated on a classical computer \cite{gottesman1997stabilizer,PhysRevA.70.052328,PhysRevA.73.022334}. The
states that can be prepared under these constraints are known as stabilizer states \cite{lami2024learning,PhysRevA.70.052328,PhysRevA.73.022334,sun2024stabilizer,gottesman1997stabilizer,PhysRevLett.128.050402}, which can manifest significant entanglement yet remain simulatable. This theorem serves as a compelling example, emphasizing that although entanglement represents a vital quantum resource, its presence alone does not suffice to render a computational problem classically hard.

In the realm of quantum computing, another crucial resource, known as non-stabilizerness or ``magic'', emerges as a another key determinant of problem complexity \cite{gu2024magicinducedcomputationalseparationentanglement}. Analogous to entanglement, non-stabilizerness has been rigorously quantified within the framework of resource theory through dedicated measures. Quantum information theory has witnessed the proposal of several such measures, underscoring the significance of non-stabilizerness in elucidating the computational capabilities of quantum systems \cite{PhysRevLett.128.050402,lami2024learning,tarabunga2024nonstabilizerness,PhysRevLett.131.180401,frau2024nonstabilizerness,PhysRevLett.112.240501,PRXQuantum.3.020333,PhysRevA.71.022316,10.21468/SciPostPhys.16.2.043}. 

This raises an intriguing question: considering that Clifford circuits effectively act as specialized disentanglers, can we seamlessly integrate them into the MPS framework to enhance MPS capabilities in the regard of entanglement entropy as we did in FAMPS \cite{Qian_2023,qian2024parent} Recent theoretical explorations have suggested that the combination of MPS with Clifford circuits can generate profoundly non-trivial quantum states \cite{lami2024quantum}. However, effectively combining MPS and Clifford formalisms remains an ongoing challenge, awaiting further investigation within the research community \cite{lami2024quantum,mello2024hybrid,lami2024learning,doi:10.1021/acs.jctc.3c00228,PhysRevA.79.022317,masotllima2024stabilizer}. 

In this Letter, we demonstrate the efficient integration of Clifford circuits into MPS with minimal modifications to the MPS algorithm. Our numerical simulations reveal that this approach maintains nearly identical computational complexity with MPS while significantly improving simulation accuracy.

{\em Clifford Circuits Augmented MPS (CAMPS)--}
DMRG is now arguably the workhorse for the accurate simulation of one-dimensional and quasi-one-dimensional quantum systems \cite{PhysRevLett.69.2863,PhysRevB.48.10345,RevModPhys.77.259}. As a variational method, the wave-function ansatz of DMRG is known as Matrix Product States (MPS) \cite{PhysRevLett.75.3537}, which is defined as
\begin{equation}
    |\text{MPS} \rangle=\sum_{\{\sigma_i\}} \text{Tr}[M_1^{\sigma_1}M_2^{\sigma_2}M_3^{\sigma_3}\cdots M_N^{\sigma_N}]|\sigma_1 \sigma_2 \sigma_3\cdots \sigma_N\rangle
\end{equation}
where $M$ is a rank-3 tensor with one physical index $\sigma_i$ (with dimension $d$) and two auxiliary indices (with dimension $D$).

Clifford circuits are sequences of quantum gates composed solely of Clifford gates. These gates include the Hadamard gate, the phase gate $S$, and the controlled-NOT (CNOT) gate. Clifford circuits are notable for their ability to generate highly entangled but efficiently simulable quantum states, as characterized by the Gottesman-Knill theorem \cite{gottesman1997stabilizer,PhysRevA.70.052328,PhysRevA.73.022334}. By acting the Clifford circuits on the physical bond of MPS, we obtain the wave function of CAMPS as
\begin{equation}
    |\text{CAMPS} \rangle=\mathcal{C}|\text{MPS}\rangle
\end{equation}
where $\mathcal{C}$ denote the Clifford circuits. A diagrammatic representation of the wave function of CAMPS is shown in Fig.~\ref{CAMPS} (a). 
In contrast to FAMPS, where disentangler positions are predetermined, as discussed below, we can embed the process of adding Clifford circuits within the MPS optimization process.

A general Hamiltonian for spin-1/2 systems with N sites can be expressed as a summation of a series of Pauli strings $P=\sigma_1 \otimes \sigma_2\cdots \otimes \sigma_N$ ($\sigma_i\in \{I, \sigma^x, \sigma^y, \sigma^z\}$)
\begin{equation}
    H=\sum_{i=1}^{m}a_i P_i
\end{equation}
where $a_i$ denote the interaction strength associated with Pauli string $P_i$, and $m$ is the total number of terms in the Hamiltonian.

In a two-site DMRG algorithm, we decompose the computation of the ground state into a series of local eigenvalue problems at sites $k,k+1$ $(k\in \{1, 2, \cdots, N-1\})$ with respect to the effective Hamiltonian
\begin{equation}
    \label{H_eff}
H_{\text{eff}}=\sum_{i=1}^{m} a_i A_{i,k-1} \otimes \sigma_{i,k} \otimes \sigma_{i,k+1} \otimes B_{i,k+2}
\end{equation}
where $A_{i,k}, B_{i,k}$ are the so-called left and right environment for $P_i$ at site $k$, $\sigma_{i,k}$ is the Pauli matrix of $P_i$ at site $k$ \cite{RevModPhys.77.259}.
Calculating the ground state of $H_{\text{eff}}$ we obtain an optimized local state $|\phi\rangle$ with $H_{\text{eff}}|\phi\rangle=E_g |\phi\rangle$. In MPS, we typically perform a Singular Value Decomposition (SVD) on $|\phi\rangle$ and truncate based on the singular values to obtain the optimized tensors: $M_k$ and $M_k+1$. However, in CAMPS, we take a different approach: first, we apply a two-qubit Clifford circuit $C$ to $|\phi\rangle$ to obtain $C|\phi\rangle$, and then we perform singular value decomposition on $C|\phi\rangle$ to minimize truncation loss. The purpose of this additional step is to transfer the stabilizer-related entropy (the entanglement entropy which can be disentangled by Clifford circuits) 
in state $|\phi\rangle$ into the Clifford circuits, thereby reducing the bond dimension needed for MPS to accurately represent the ground state. Given that $C H_{\text{eff}}C^\dagger C|\phi\rangle=E_g C|\phi\rangle$, after obtaining the optimized tensors $M_k, M_k+1$, we then need to perform a transformation to the original Hamiltonian accordingly 
\begin{equation}
    \label{H_prime}
    \begin{split}
    H^\prime =&CHC^\dagger=\sum_{i=1}^{m}a_i CP_iC^\dagger\\
            =&\sum_{i=1}^{m}a_i \sigma_1 \otimes \cdots \sigma_{k-1} C\sigma_{i,k} \otimes \sigma_{i,k+1} C^\dagger \sigma_{i,k+2} \otimes \cdots \sigma_{N}\\
            =&\sum_{i=1}^{m}a^{\prime}_i \sigma_1 \otimes \cdots \sigma_{k-1} \sigma^{'}_{i,k} \otimes \sigma^{'}_{i,k+1} \sigma_{i,k+2} \otimes \cdots \sigma_{N}
    \end{split}
\end{equation}
where the Clifford circuit $C$ only acts on sites $k$ and $k+1$. In the last line, we take the advantage of the fact that the Clifford circuits preserve the Pauli string form when acting on a Pauli string  $CPC^\dagger=P^{\prime}$, showcasing the advantage of using Clifford circuits as disentanglers over employing a general unitary matrix. For general disentanglers, the summation over $m$ grows exponentially with the number of layers of disentanglers. However, by utilizing Clifford circuits as disentanglers, one can add infinite layers of disentanglers without worrying about the increase of the computational cost. After transforming the Hamiltonian $H$ to $H^\prime$, the subsequent steps are the same as in the process in DMRG, updating the left environment tensor $A_{i,k}$, moving to the next sites $k+1, k+2$, and iteratively repeating this process.

Now, the primary issue is to identify the optimal Clifford circuit to minimize truncation loss for $C|\phi\rangle$, which is equivalent to minimize the discarded singular values. With a total of 720 two-qubit Clifford circuits (excluding phase redundancy, as they do not affect singular values) \cite{10.1063/1.4903507,PhysRevB.100.134306,PhysRevA.87.030301}, calculating the singular values for all $C|\phi\rangle$ allows for the determination of the optimal Clifford circuit. The illustration of this process is provided in Fig.~\ref{CAMPS} (b).

\begin{figure}[t]
    \includegraphics[width=40mm]{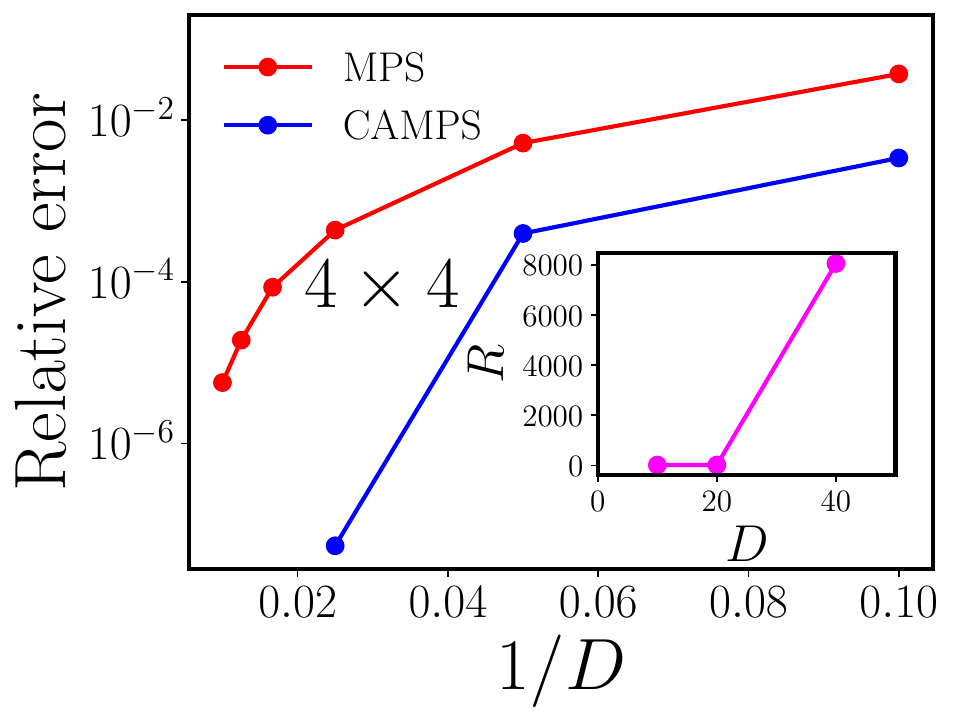}
    \includegraphics[width=40mm]{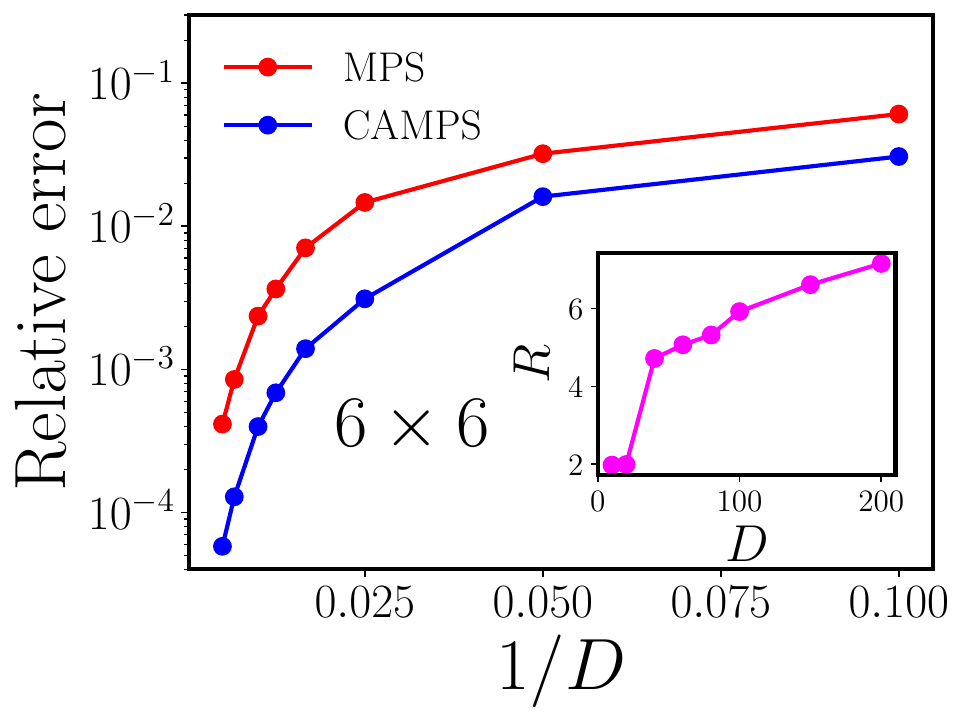}
    \includegraphics[width=40mm]{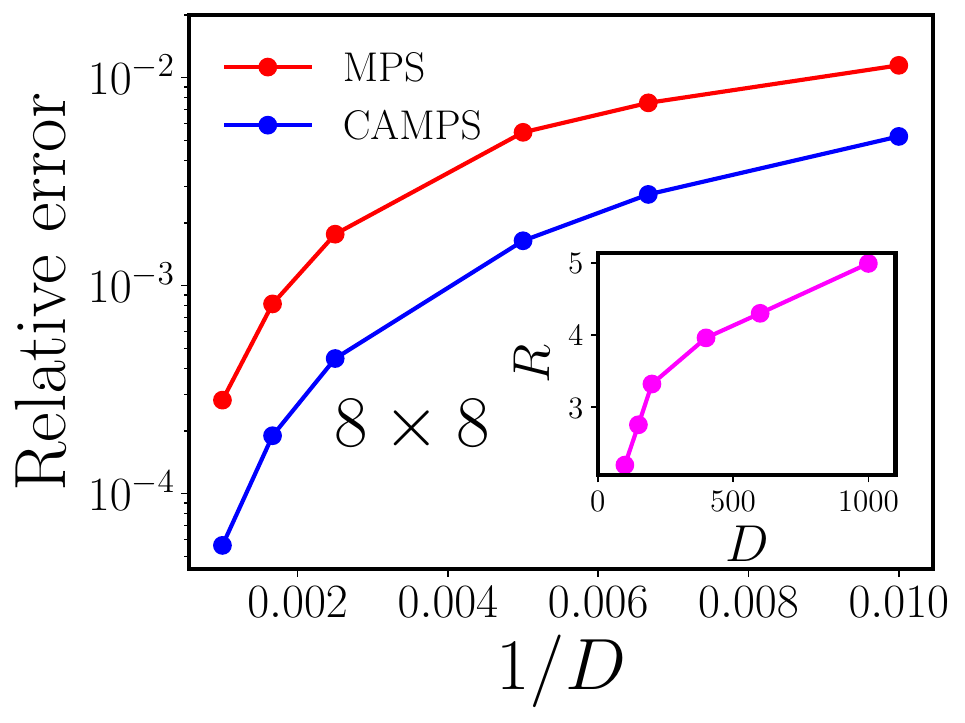}
    \includegraphics[width=40mm]{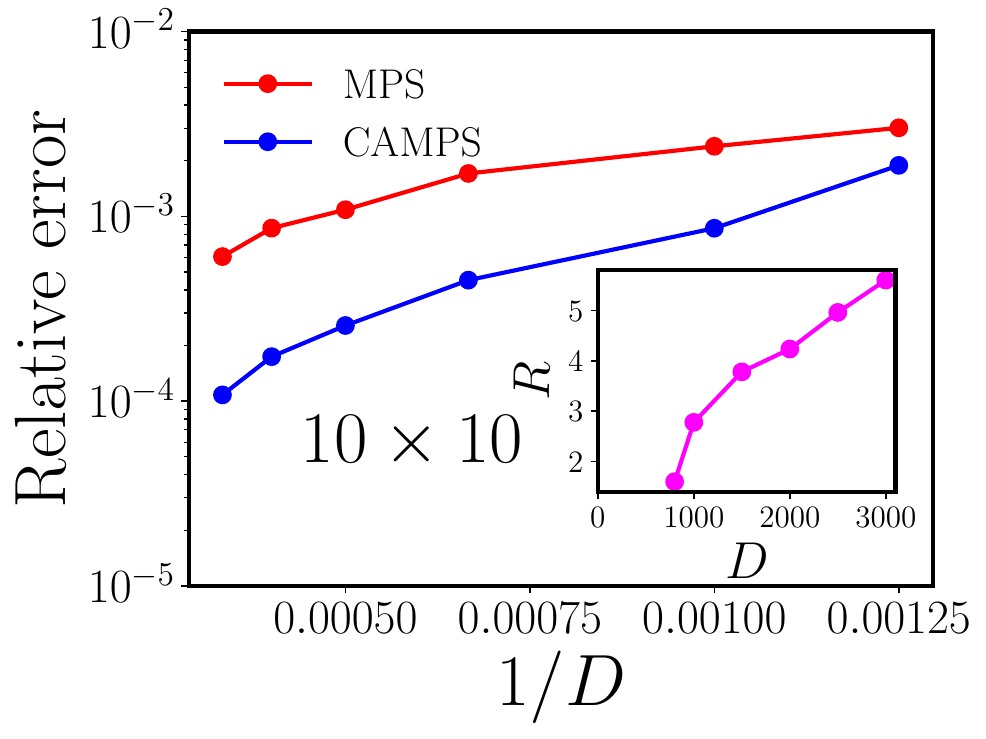}
       \caption{Relative error of the ground state energy for CAMPS and MPS of the $J_1-J_2$ Heisenberg model at $J_2=0$ as a function of bond-dimension $D$ for different lattice sizes $L\times L$. Open boundary conditions are considered in the simulations. The inset shows the ratio $R$ of the relative error of MPS and CAMPS. We observe that the ratio increases with the bond dimension, indicating that the improvement amplifies with larger bond dimensions.}
       \label{Hei}
\end{figure}

{\em Simulations on 2D $J_1-J_2$ Heisenberg model.--}
Here, we test our approach on the 2D $J_1-J_2$ Heisenberg model. The Hamiltonian of the model is defined as
\begin{equation}
	H = J_1\sum_{\langle i,j \rangle}S_{i} \cdot S_{j}+J_2\sum_{\langle\langle i,j \rangle\rangle}S_{i} \cdot S_{j}
\end{equation}
where $S_i$ is the spin-1/2 operator on site $i$, and the summations are taken over nearest-neighbor ($\langle i,j \rangle$) and next-nearest-neighbor ($\langle \langle i,k \rangle \rangle$) pairs. To implement MPS for 2D systems, we use the standard snake-like mapping to convert the 2D lattice into an effective 1D structure. Specifically, we convert the 2D coordinates $(x, y)$ to a 1D index $(x \times L_y + y)$ with $L_y$ the width of the system.

\begin{figure}[t]
    \includegraphics[width=40mm]{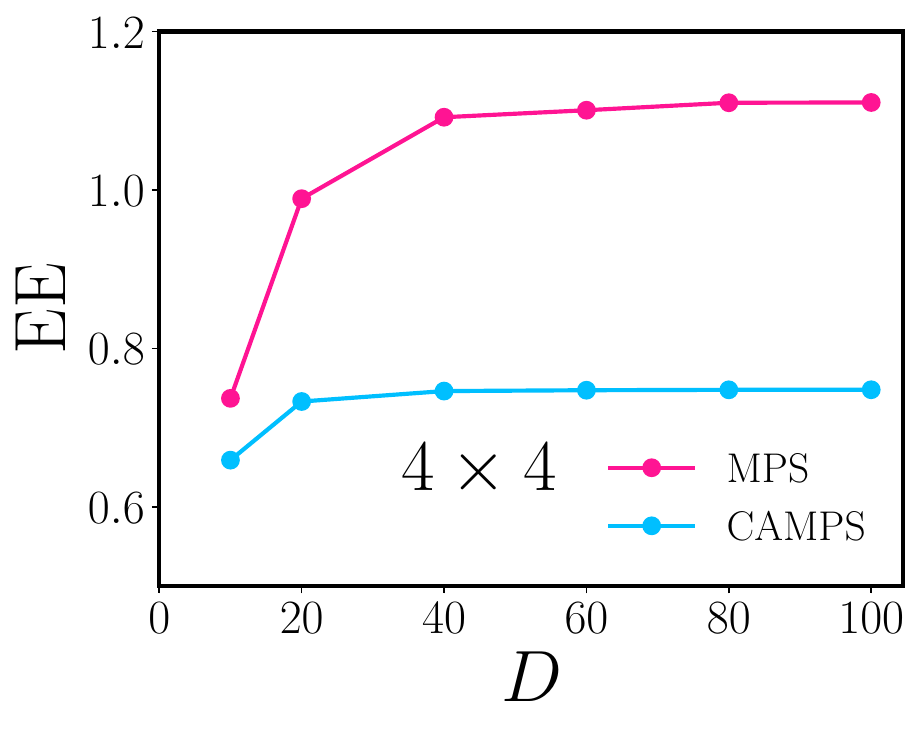}
    \includegraphics[width=40mm]{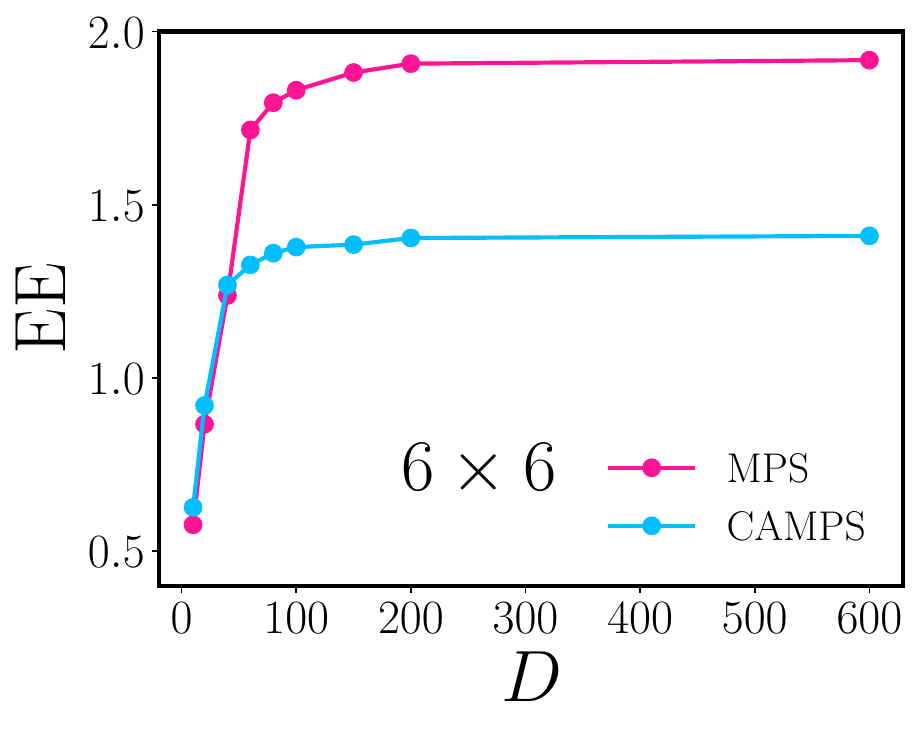}
    \includegraphics[width=40mm]{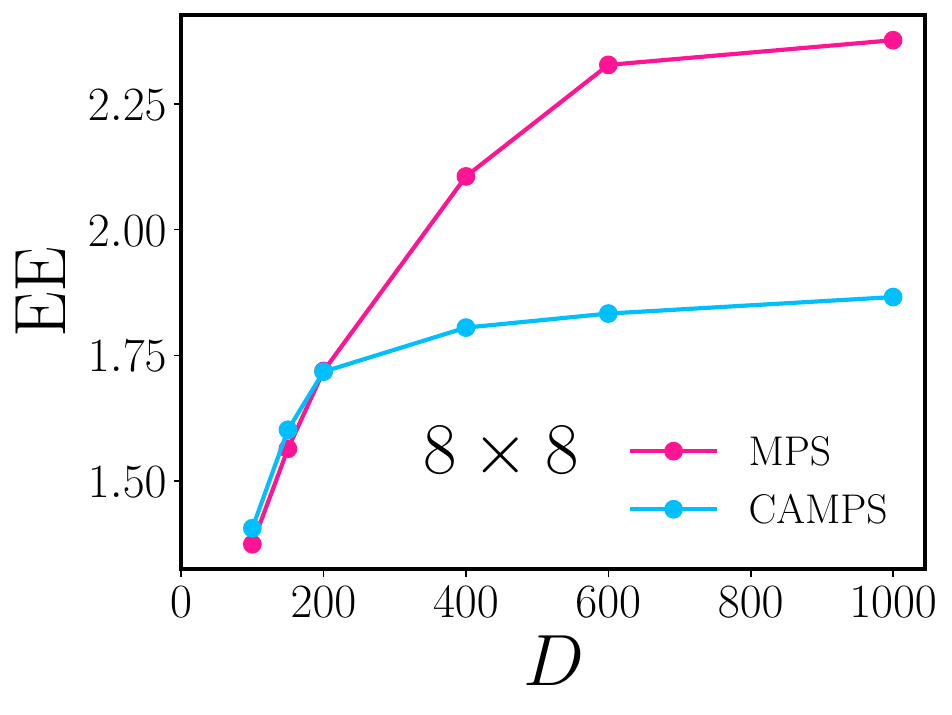}
    \includegraphics[width=40mm]{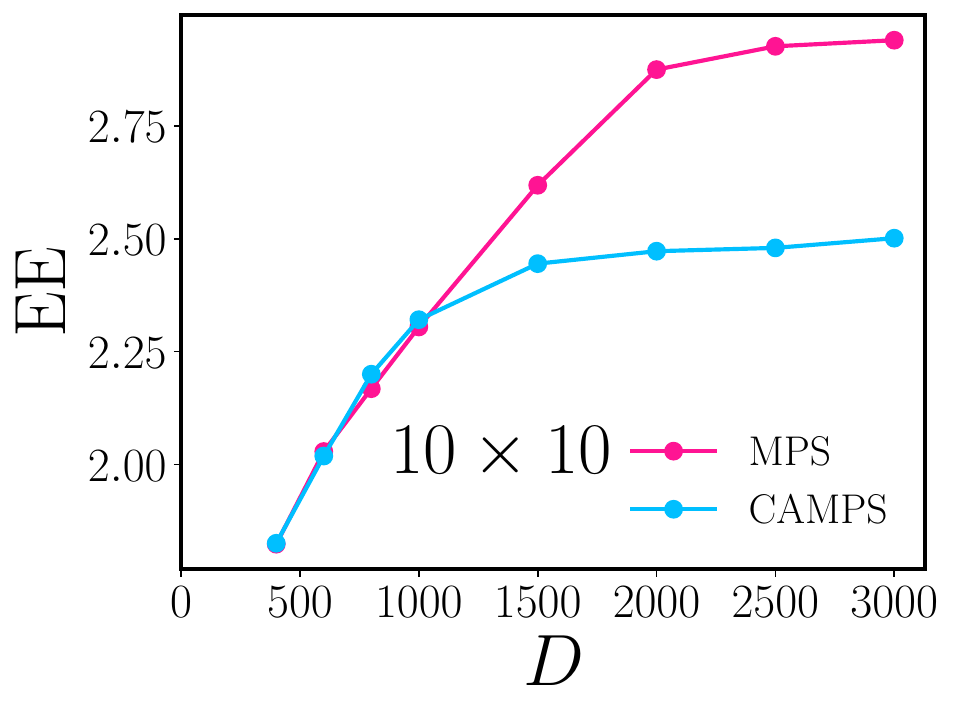}
       \caption{
        Entanglement entropy (EE) at the center bond in the MPS part for CAMPS and pure MPS of the $J_1-J_2$ Heisenberg model at $J_2=0$ as a function of bond-dimension $D$ for various lattice sizes $L\times L$. Open boundary conditions are considered in the simulations. The entanglement entropy in the MPS part of CAMPS is smaller compared to a pure MPS calculation as expected. Interestingly, we observe a critical bond-dimension threshold, before which the entanglement entropy in MPS for both pure MPS and CAMPS calculations remains nearly identical. Beyond this critical bond-dimension, the entanglement entropy in the MPS part for CAMPS saturates rapidly, whereas in the case of MPS, it continues to increase with bond dimension.}
       \label{Hei_EE}
\end{figure}

First we consider the case when $J_2=0$, where we have numerically exact Quantum Monte Carlo results as a reference \cite{PhysRevB.103.235155,PhysRevE.66.046701}. In Fig.~\ref{Hei} we show the relative error of ground state energy for lattice sizes $4\times4, 6\times6, 8\times 8$, and $10\times 10$ with open boundary conditions. For all the system sizes, we are able to achieve a simulation accuracy of the order of $10^{-4}$ with CAMPS. For the $10\times 10$ system, the relative error is reduced by a factor of $5$ with $D=3000$, and this factor further increases with bond-dimension $D$. To further show that our approach can reduce the entanglement in the MPS, we also calculate the entanglement entropy at the center bond encoded in the MPS part for CAMPS and in pure MPS. The results are show in Fig.~\ref{Hei_EE}. The entanglement entropy in the MPS part of CAMPS is smaller compared to a pure MPS calculation as expected. Interestingly, we observe a critical bond-dimension threshold, before which the entanglement entropy in MPS for both pure MPS and CAMPS calculations remains nearly identical. Beyond this critical bond-dimension, the entanglement entropy in the MPS part for CAMPS saturates rapidly, whereas in the case of MPS, it continues to increase with bond dimension. This is in agreement with previously studies \cite{frau2024nonstabilizerness} which shows that the non-stabilizerness in MPS converges faster than entanglement entropy. As the stabilizer part of a quantum state is encapsulated by the additional Clifford circuits in CAMPS, CAMPS is expected to converge rapidly once the non-stabilizerness of the state is converged. This highlights the advantages of CAMPS over pure MPS calculations, as it allows the MPS component to deal solely with the non-stabilizerness.

\begin{figure}[t]
    \includegraphics[width=40mm]{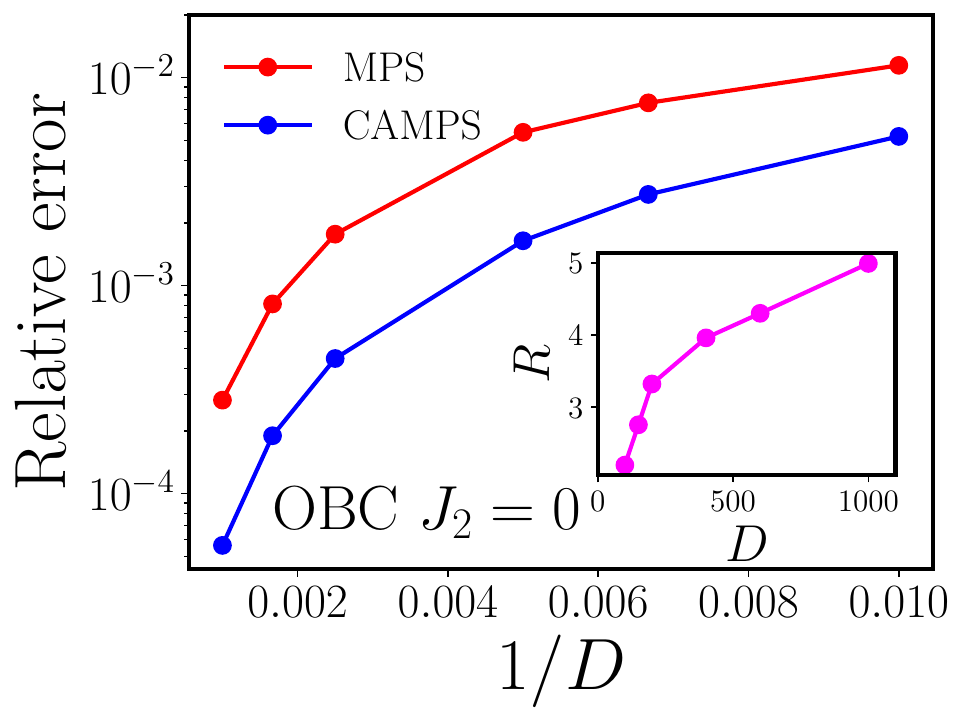}
    \includegraphics[width=40mm]{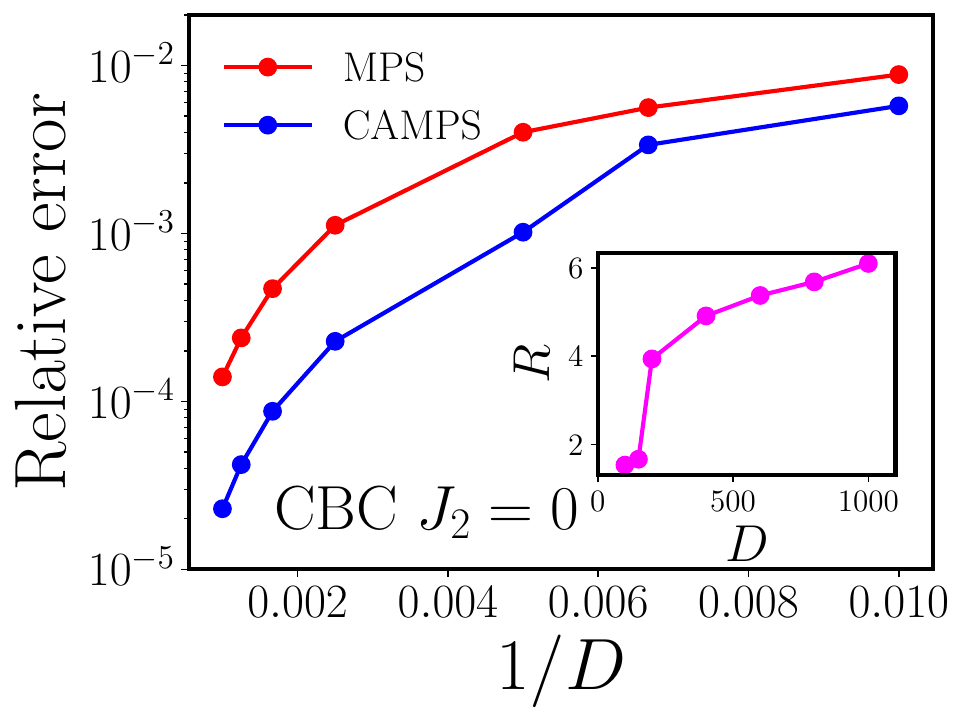}
    \includegraphics[width=40mm]{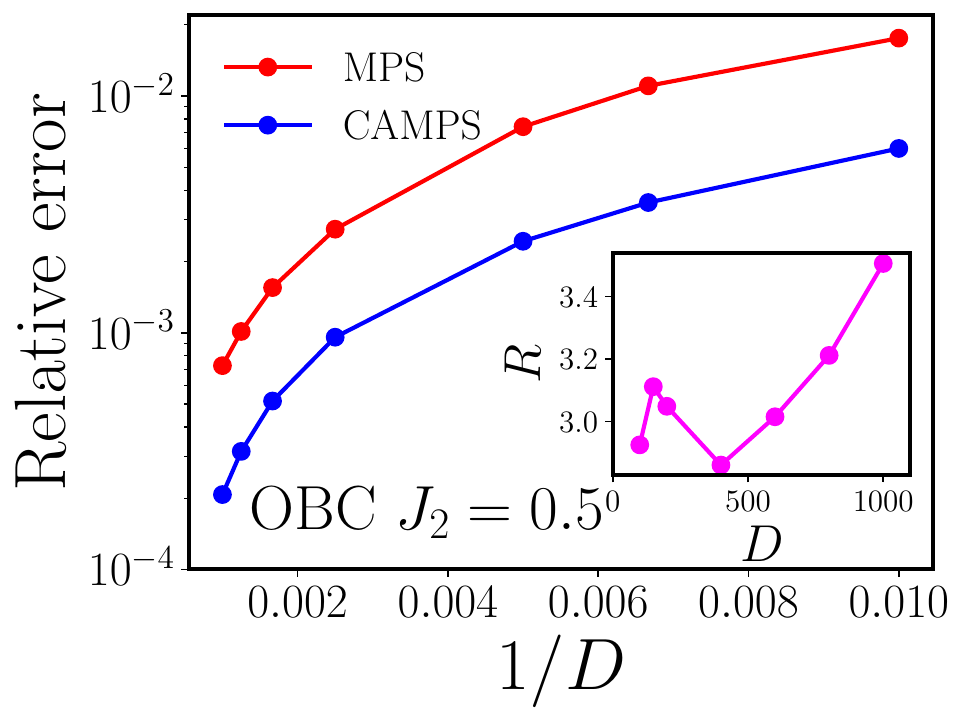}
    \includegraphics[width=40mm]{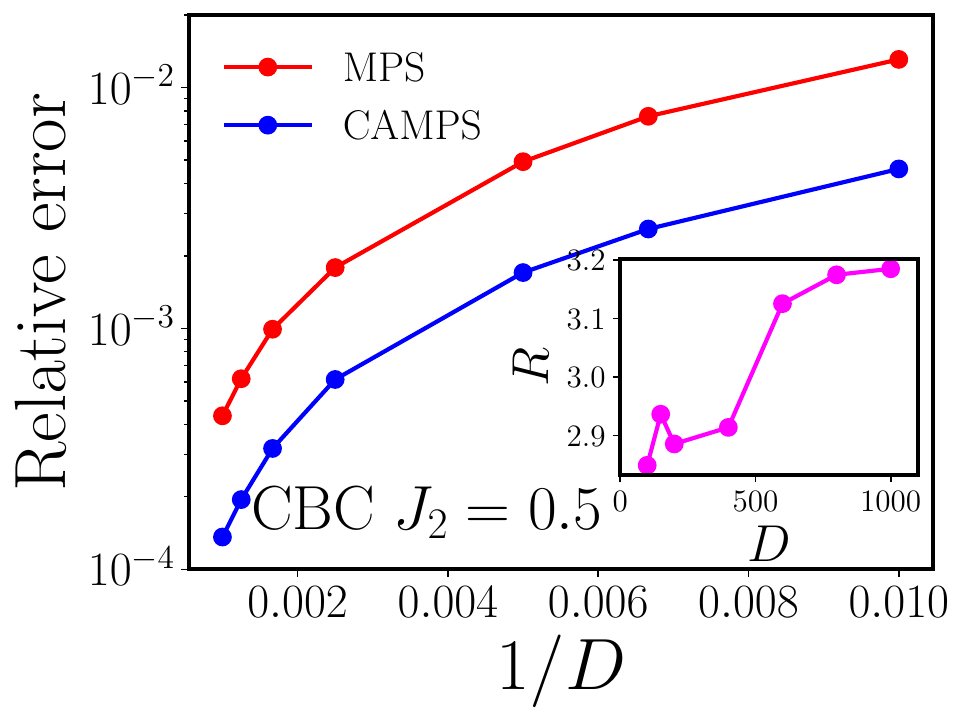}
       \caption{Relative error of the ground state energy for CAMPS and MPS of the $J_1-J_2$ Heisenberg model at $J_2=0$ and $J_2=0.5$ as a function of bond-dimension $D$ for a $8\times 8$ lattice. Open boundary conditions (OBC) and Cylinder boundary conditions (CBC) are considered for comparison. The improvement of CAMPS over MPS remains consistently robust across different $J_2$ values and boundary conditions. In our test cases,  achieving a precision of $10^{-4}$ with a bond dimension of $D \approx 1000$ is readily attainable with CAMPS. The inset shows the ratio $R$ of the relative error of MPS and CAMPS. The ratio also demonstrates an increasing trend with bond dimension.}
       \label{Hei_C}
\end{figure}

We then test our approach across different parameters of the model and under different boundary conditions. The results are shown in Fig.~\ref{Hei_C}. In Fig.~\ref{Hei_C}, we show the results of the relative error of the ground state energy for a $8\times 8$ lattice under different boundary conditions at $J_2=0$ and $J_2=0.5$. The reference energy is taken from an MPS calculation with $D=10000$. The improvement of CAMPS over MPS remains consistently robust across different $J_2$ values and boundary conditions. In our test cases, achieving a precision of $10^{-4}$ with a bond dimension of $D \approx 1000$ is readily attainable with CAMPS.

{\em Discussion.}
The primary distinction of CAMPS compared to MPS lies in the additional step of identifying the optimal Clifford circuit to mitigate truncation loss. In our practical simulations, this supplementary process requires a time of the same order as finding the ground state of $H_{\text{eff}}$. However, the time required for finding the optimal Clifford circuit can be expedited by either sampling the two-qubit Clifford group or exhaustively parallelizing. For the $10 \times 10$ Heisenberg model under OBC, the ratio of calculation time for CAMPS and MPS is about $1.2$, which becomes closer to $1$ with the increase of bond dimension. In practical MPS simulation, one can add different $A_{i,k}$ or $B_{i,k}$ together for some Pauli strings sharing the same interaction part to reduce the computational cost, which can reduce the summation of $m$ (equivalent to the bond-dimension of the Matrix Product Operator (MPO) for $H_{\text{eff}}$\cite{RevModPhys.77.259}) in the effective Hamiltonian $H_{\text{eff}}$ in Eq.~\ref{H_eff} from $O(N)$ to $O(\sqrt{N})$ for a 2D system with local interactions. This is the true case in our practical simulations. One may expect, in CAMPS, the final $H^{\prime}$ would be very long-ranged, and the summation of $m$ is of the order of $O(N)$. However, in our calculations on the $J_1-J_2$ Heisenberg model, we find that the interactions in $H^{\prime}$ remain relatively local \footnote{{The average interaction length is defined as the average number of sites that the interactions span in the 1D setup. For the models studied in this work, the average interaction length is approximately the same for MPS and CAMPS.}}, resulting in an average length of the effective terms in $H^{\prime}$ that is nearly identical to that in a pure MPS calculation.

The optimization framework illustrated in Fig.~\ref{CAMPS} (b) can be readily extended to other Tensor Network States, such as PEPS. One can also use the symmetry preserved Clifford circuits \cite{PRXQuantum.4.040331} to implement symmetries in CAMPS, thereby enhancing efficiency. Another important issue is that the local optimization of Clifford circuits within our framework may lead to being trapped in local minima \footnote{Empirically, local minima is encountered in calculations with small bond dimensions. So it is better to not apply Clifford circuits when bond dimension is small.}. Employing more sophisticated optimization techniques could yield optimal Clifford circuits for CAMPS more effectively.

{\em Conclusion and Perspective.}
In this study, we introduce a new Tensor Network ansatz, CAMPS, in which Matrix Product States (MPS) is augmented by additional Clifford circuits. Crucially, we devise an efficient method to seamlessly integrate Clifford circuits into MPS with minimal modifications to the existing algorithm. Notably, our approach maintains the same computational complexity as MPS, with only slight additional computational cost. We demonstrate the efficiency of our method by benchmarking it on the 2D $J_1-J_2$ Heisenberg model. Our results exhibit significant improvements over pure MPS calculations, underscoring the effectiveness and potential of our approach in unraveling the mysteries of 2D quantum systems. Furthermore, we anticipate that our framework can be readily extended to other tensor network methods. The idea in our work could also stimulate the combination of Clifford circuits and numerical simulation frameworks other then tensor network.
We only test the approach in spin models in this work, but it will be interesting to incorporate fermionic degrees of freedom in CAMPS in the future.


\textbf{Acknowledgments:} The calculation in this work is carried out with TensorKit \cite{foot7}. The computation in this paper were run on the Siyuan-1 cluster supported by the Center for High Performance Computing at Shanghai Jiao Tong University. MQ acknowledges the support from the National Natural Science Foundation of China (Grant No. 12274290), the Innovation Program for Quantum Science and Technology (2021ZD0301902), and the sponsorship from Yangyang Development Fund.

\bibliography{main}

\end{document}